\documentclass[12pt]{article}

\begin{document}

\def\CA{{\cal A}}
\def\CB{{\cal B}}
\def\CC{{\cal C}}
\def\CD{{\cal D}}
\def\CE{{\cal E}}
\def\CF{{\cal F}}
\def\CG{{\cal G}}
\def\CH{{\cal H}}
\def\CI{{\cal I}}
\def\CJ{{\cal J}}
\def\CK{{\cal K}}
\def\CL{{\cal L}}
\def\CM{{\cal M}}
\def\CN{{\cal N}}
\def\CO{{\cal O}}
\def\CP{{\cal P}}
\def\CQ{{\cal Q}}
\def\CR{{\cal R}}
\def\CS{{\cal S}}
\def\CT{{\cal T}}
\def\CU{{\cal U}}
\def\CV{{\cal V}}
\def\CW{{\cal W}}
\def\CX{{\cal X}}
\def\CY{{\cal Y}}
\def\CZ{{\cal Z}}

\newcommand{\btzm}[0]{BTZ$_{\rm M}$}
\newcommand{\todo}[1]{{\em \small {#1}}\marginpar{$\Longleftarrow$}}
\newcommand{\labell}[1]{\label{#1}\qquad_{#1}} 
\newcommand{\ads}[1]{{\rm AdS}_{#1}}
\newcommand{\SL}[0]{{\rm SL}(2,\IR)}
\newcommand{\cosm}[0]{R}
\newcommand{\tL}[0]{\bar{L}}
\newcommand{\hdim}[0]{\bar{h}}
\newcommand{\bw}[0]{\bar{w}}
\newcommand{\bz}[0]{\bar{z}}
\newcommand{\be}{\begin{equation}}
\newcommand{\ee}{\end{equation}}
\newcommand{\lp}{\lambda_+}
\newcommand{\bx}{ {\bf x}}
\newcommand{\bk}{{\bf k}}
\newcommand{\bb}{{\bf b}}
\newcommand{\BB}{{\bf B}}
\newcommand{\tp}{\tilde{\phi}}
\hyphenation{Min-kow-ski}

\def\ie{{\it i.e.}}
\def\eg{{\it e.g.}}
\def\cf{{\it c.f.}}
\def\etal{{\it et.al.}}
\def\etc{{\it etc.}}

\def\apr{\alpha'}
\def\str{{str}}
\def\lstr{\ell_\str}
\def\gstr{g_\str}
\def\Mstr{M_\str}
\def\lpl{\ell_{pl}}
\def\Mpl{M_{pl}}
\def\varep{\varepsilon}
\def\del{\nabla}
\def\grad{\nabla}
\def\tr{\hbox{tr}}
\def\perp{\bot}
\def\half{\frac{1}{2}}
\def\p{\partial}

\renewcommand{\thepage}{\arabic{page}}
\setcounter{page}{1}

\rightline{HUTP-00/A007, CITUSC/00-013}
\rightline{CALT68-2265, hep-th/0003147}
\vskip 1cm
\centerline{\Large \bf Consistency Conditions for Holographic Duality}
\vskip 1cm

\renewcommand{\thefootnote}{\fnsymbol{footnote}}
\centerline{{\bf Vijay
Balasubramanian${}^{1}$\footnote{vijayb@pauli.harvard.edu},
Eric Gimon${}^{2}$\footnote{egimon@theory.caltech,edu},
 and
Djordje Minic${}^{3}$\footnote{minic@physics.usc.edu}}} 	
\vskip .5cm
\centerline{${}^1$\it Jefferson Laboratory of Physics, Harvard
University,}
\centerline{\it Cambridge, MA 02138, USA.}
\vskip .5cm
\centerline{${}^2$ \it CIT-USC Center for Theoretical Physics} 
\centerline{\it California Institute of Technology, Pasadena, CA
91125.}
\vskip .5cm 
\centerline{${}^3$ \it CIT-USC Center for Theoretical Physics}
\centerline{\it Department of Physics and Astronomy, University of
Southern California}
\centerline{\it Los Angeles, CA 90089-0484, USA.}

\setcounter{footnote}{0}
\renewcommand{\thefootnote}{\arabic{footnote}}

\begin{abstract} 

We show that if the beta functions of a field theory are given by the
gradient of a certain potential on the space of couplings, a gravitational
background in one more dimension can express the renormalization group (RG)
flow of the theory.  The field theory beta functions and the gradient flow
constraint together reconstruct the second order spacetime equations of
motion.  The RG equation reduces to the conventional gravitational
computation of the spacetime quasilocal stress tensor, and a c-theorem
holds true as a consequence of the Raychaudhuri equation.  Conversely,
under certain conditions, if the RG evolution of a field theory possesses a
monotonic c-function, the flow of couplings can be expressed in terms of a
higher dimensional gravitational background.

\end{abstract}

\newpage

{\leftline {\large {\bf Introduction}}}

The holographic principle states that the degrees of freedom
describing quantum gravity in some volume can be encoded at fixed
density on a ``screen" or surface that bounds that
volume~\cite{holog,bousso}.  A particularly simple realization of this
principle appears to occur for asymptotically anti-de Sitter (AdS)
spaces where $(d+1)$-dimensional spacetime dynamics is conjectured to
be encoded in a local quantum field theory (QFT) on the timelike,
$d$-dimensional AdS boundary~\cite{juanads}.

In this correspondence, phenomena occuring closer to the AdS boundary are
related to local, ultraviolet physics in the field theory, while infrared
and non-local data encode the deep interior of the space~\cite{susswitt,
amanda,bklt,bdhm,fatcats,swedes,easylife}.  This suggests that the
semiclassical structure called spacetime can sometimes be generated from a
QFT via renormalization group (RG) flow.\footnote{In fact, it is known that the
string field theory equations of motion are closely related to the
Wilsonian RG flow of the string sigma model~\cite{banksmart}.}

Here, we show that if the beta functions of a field theory in $4$
dimensions are given by the gradient of a certain potential on the space of
couplings, the RG flow of the theory admits a description in terms of $5$
dimensional gravity coupled to scalar fields.  For such flows, the second
order spacetime equations of motion can be reconstructed from the field
theory beta functions and the gradient flow constraint.  The field theory
RG equation is realized as the conventional gravitational computation of
the trace of the quasilocal stress tensor.  A c-theorem is satisfied by the
RG flow as a consequence of the Raychaudhuri equation for the $5$
dimensional gravitational background.  Conversely, if a QFT has a
c-function that evolves monotonically down an RG trajectory, under certain 
conditions the flow of couplings may be expressed in terms of a $5$
dimensional gravitational background.

\vskip0.1in
{\leftline {\large {\bf Gradient beta functions}}}

Consider a $4$-dimensional QFT on a Ricci-flat manifold with an
interaction Lagrangian:
\begin{equation}
S_{int} = \int d^4x \, \phi^I \, O_I 
\, .
\label{intlag}
\end{equation}
On a flat manifold
there is no conformal anomaly, and the RG equation is simply given by
the conformal Ward identity for the trace of the stress tensor:
\begin{equation}
\langle T \rangle \equiv 
\langle T^i_i 
\rangle \propto {d\Gamma \over d \log \lambda}
= \beta^I {{\partial\Gamma}\over {\partial\phi^I}} 
\, .
\label{trace}
\end{equation}
$\Gamma$ is quantum effective action of the field theory, the beta
functions are the scale derivatives of the couplings $\beta^I =
{{d\phi^I}\over {d \log \lambda}}$, and $\lambda = \Lambda/\Lambda_0$
in terms of an energy cutoff $\Lambda$ and a reference scale
$\Lambda_0$.

Now consider a QFT in which the beta fuctions can be derived as
gradients of a particular potential on the couplings: 
\begin{equation} 
\beta^I = - G^{IJ}(\phi) {\partial \over \partial \phi^J} \, \log({a \, \langle T
\rangle + U(\phi)}) 
\label{betaT}
\end{equation}
$G^{IJ}$ is a positive, symmetric function of the couplings $\phi^I$, $a$
is a constant of length dimension 4, and $U(\phi)$ is a function of
couplings that will be related to the norm of the beta functions.
We will show that near a conformal point the inverse metric 
$G_{IJ}$ must be related to the normalization of the 2-point
function $\langle O_I O_J \rangle$.  At a conformal point the beta
functions vanish.  Then (\ref{betaT}) implies that $\partial (\langle T
\rangle + U)\over \partial \phi^I $ also vanishes, so that $(\langle T
\rangle + U)$ is extremized as a function of the couplings.

The QFT RG flow can be related to the equations of motion of 5d gravity if,
in addition, the norm of the beta functions is:
\begin{equation}
{1\over 4}\,  G_{IJ} \,  \beta^I \, \beta^J =
1 - {V(\phi) \over (a \, \langle T \rangle + U(\phi))^2 }  \geq 0
\, .
\label{potential}
\end{equation}
$V(\phi)$ will be related to a c-function for the RG flow. The ultraviolet
limit of the theory, if it exists, is conformal.  In this limit, $\langle T
\rangle \rightarrow 0$, and we can choose the normalization $U(\phi)
\rightarrow 1$.  Furthermore, at {\it any} conformal point, the vanishing
of the beta functions requires that $V(\phi) \rightarrow \langle T \rangle
+ U(\phi)$, implying that $V(\phi)$ is extremized.  We will show that the
first order RG flow of the $d$-dimensional QFT can be represented as the
second order equations of motion of $5$-dimensional gravity coupled to
scalars $\phi^I$, with a potential $V(\phi)$ and a ``boundary'' cosmological
constant related to $U(\phi)$.

At first glance, there are three unknowns in (\ref{betaT}) and
(\ref{potential}): the metric $G^{IJ}$ and the potentials $U(\phi)$ and
$V(\phi)$.  Since there are only two equations, this suggests that {\em
any} RG flow is expressible in this manner by a suitable choice of metric
and potentials.  Near a conformal point, we will show that these quantities
relate to other dynamical data such as correlation functions.  But away
from a conformal point, despite the constraints implied by the positivity
and symmetry of the metric, (\ref{betaT}) and (\ref{potential}) seem to be
rather weak restrictions.  However, we will argue that any flow that can be
written as (\ref{betaT}) and (\ref{potential}) possesses a c-function
related to the potentials $U$ and $V$, provided a certain positive energy
condition is satisfied.  The inequality in (\ref{potential}) will be
equivalent to a statement that field theories obeying (\ref{betaT}) have a
c-function that decreases monotonically during RG flow.  So we will be
forced to conclude either that a surprisingly general class of field
theories possesses a monotonic c-function, or that the conditions
(\ref{betaT}) and (\ref{potential}) are much more restrictive than they
appear.

\vskip0.1in
{\leftline {\bf {\large Gravitational description}}}

Einstein gravity coupled to scalars on a 5-dimensional manifold $\CM$ 
has an action
\begin{eqnarray}
S_{h} &=& {1 \over 16\pi G_N}  \, \int_{\CM} d^{4}x \,  dr \, \sqrt{-h}
\, \left[ -R_h + {1 \over   2} g_{IJ} \, h^{\mu\nu} \, (\partial_\mu
\alpha^I)(\partial_\nu 
\alpha^J) - {12 \over \ell^2} \, v(\alpha) \right]   \nonumber
\\
&-&{1\over 8\pi G_N} \, \int_{\partial \CM} d^{4}x \, \sqrt{\gamma} \,
\left[\Theta + L_{c.t.} \right] 
\label{bulkact}
\end{eqnarray}
$R_h$ is the Ricci scalar of the spacetime metric $h_{\mu\nu}$, $g_{IJ}$ is
the metric on the space of scalars $\alpha^I$ and $\ell$ is a length scale.
By choice, the potential $v$ has negative extrema, with at least one
extremum at $v = -1$.  Placing the scalars at these points induces a
negative cosmological constant in the space.  We will be interested in
solutions whose potentials approach $v = -1$ as $r \rightarrow \infty$.
The boundary extrinsic curvature $\theta$ makes the equations of motion
well-defined, and $L_{{\rm c.t.}}$ is a counterterm Lagrangian constructed
from intrinsic invariants of the induced metric $\gamma_{ij}$ on the
spacetime boundary.  When the scalars are at the $v(\alpha) = -1$ extremum,
the gravitational part of the action can be rendered finite by setting $
L_{{\rm c.t.}}= {3 \over \ell} \left( 1 - {\ell^2 \over 12} R_\gamma
\right)$~\cite{hologanom,adsstress}.  We require a counterterm scheme that
yields a finite gravitational action for any solution that asymptotically
approaches an extremum of the scalar potential.  One such scheme is
\begin{equation}
L_{{\rm c.t.}}= {3 \over \ell} \left( u(\alpha) - {\ell^2 \,
\over 12\, u(\alpha)} R_\gamma \right)  \, ,
\label{counter}
\end{equation}
subject to the requirement that $u(\alpha)^2 \rightarrow -v(\alpha)$ when
the $\alpha^I$ approach any extremum of $v$.\footnote{The arguments of
$u(\alpha)$ are the boundary values of the scalars.  Also
see~\cite{othercounter} for discussions for boundary counterterms in
theories with scalars.} In  effect, $u(\alpha)$
serves as a cosmological constant on the boundary of the space. Similarly,
counterterms may be added to cancel divergences in the total action for the
scalar fields~\cite{gordon}.\footnote{As we will discuss later, there is a
large scheme dependence in this counterterm prescription, in parallel with
scheme dependences in field theory RG flows.}

The most general $4$-dimensional Poincar\'e invariant solution of this
action can be put in the form 
\begin{equation}
ds^2 = e^{2A(r)} \, \eta_{ij} \, dx^i dx^j + dr^2
\, ,
\label{warpmet}
\end{equation}
with $\eta_{ij}$ the flat metric in $4$ dimensions and the scalars
$\phi$ chosen as functions of $r$ only.  The second order equations of
motion that follow from varying  (\ref{bulkact}) with respect to the
metric and  requiring a solution of the form (\ref{warpmet})
are~\cite{MIT}:
\begin{eqnarray}
{d^2 A \over d r^2} &=&
-{1 \over 6}  \, g_{IJ}\, {d \alpha^I \over dr}\, {d \alpha^J \over dr}
\, , \label{firsteom} \\
\left({d A \over d r}\right)^2 &=&
- {1 \over \ell^2} \, v(\alpha)
+ {1 \over 24} \, 
g_{IJ} \,
{d \alpha^I \over d r} \,
{d \alpha^J \over d r} 
\, . 
\label{secondeom}
\end{eqnarray}
The second equation is simply the statement of $r$-reparametrization
invariance of the action (\ref{bulkact}) -- i.e., it is the
Hamiltonian constraint.  The scalar equations of motion are 
\begin{equation}
g_{KL} \, {d^2\alpha^L \over d r^2} + 4 \, g_{KL} \, {d \alpha^L \over
dr} \, {d A \over d r} = 
{12 \over \ell^2} \, {\partial v \over \partial \alpha^K} +
{1 \over 2} {\partial g_{IJ} \over \partial \alpha^K} \, {d \alpha^I
\over dr} \, {d \alpha^J \over dr}  -
{\partial g_{KL} \over \partial \alpha^I} \, 
{d\alpha^I \over dr} \, {d\alpha^L \over dr} \, .
\label{scaleom}
\end{equation}

So long as $d\alpha^I / dr \neq 0$, we can treat $A' \equiv d A /d r$ and
$\alpha^{I'} \equiv d \alpha^I /d r$ as functions of $\alpha$, to write
(\ref{firsteom}) as\footnote{The analysis below will always hold piecewise
in domains where $d\alpha^I/dr \neq 0$.  When the scalars are at an
extremum of $v$, $\alpha^{I'} = 0$ and $A'$ is a constant.  This will
correspond to a conformal point in the field theory.}
\begin{eqnarray}
{\partial A' \over \partial \alpha^I} \, {d\alpha^I \over d r} &=&
- {1 \over 6} g_{IJ} \, {d\alpha^I \over dr} \, {d\alpha^J \over dr}
\,
.
\label{3eom}
\end{eqnarray}
Solutions to (\ref{3eom}) are obtained by setting
\begin{equation}
{d \alpha^I \over dr} = -6 \, g^{IJ} \, {\partial A' \over \partial
\alpha^J}
.
\label{alphasoln}
\end{equation}
Computing $A'(\alpha)$ from (\ref{alphasoln}) and (\ref{secondeom}),
it is easy to show that the scalar equations (\ref{scaleom}) are
automatically satisfied.\footnote{To show this, differentiate
(\ref{alphasoln}) with respect to $r$ and use the $\alpha^I$
derivative of (\ref{secondeom}).}  Thus, starting from (\ref{alphasoln}) 
and (\ref{secondeom}) one can derive the complete equations of motions 
for a five-dimensional lagrangian of the form (\ref{bulkact}). 

Integrating the solutions for $\phi^{I'}$ and $A'$ yields a trajectory
$\alpha^I(r)$ in the N-dimensional space of scalar fields.  The $N$
first-order equations (\ref{3eom}) and the the N-dimensional first-order
equation (\ref{secondeom}) produce $2N$ integration constants~\cite{MIT}.
Along with the specification of the integration bound $r_0$ and $A(r_0)$,
these constants reproduce the $2N + 2$ expected initial conditions of the
scalar and gravity equations of motion~\cite{MIT}.

Now consider the total action as a functional of the induced
``boundary'' metric $\gamma_{ij}$ on a surface of fixed $r$.  Following
Brown and York~\cite{brownyork}, and including the counterterms
(\ref{counter})~\cite{adsstress}, the quasilocal stress tensor of the
spacetime on a fixed-$r$ surface is the response of the action
to variations of $\gamma_{ij}$:
\begin{eqnarray}
\tau_{ij} &=& {2 \over \sqrt{- \gamma}} \, {\delta S_h \over \delta
\gamma^{ij}} 
\label{HJ1}
\\
&=&
{1 \over 8\pi G_N} \,
\left[ 
\theta_{ij} - \theta \, \gamma_{ij} - {3 \, u(\alpha) \over \ell}
\, \gamma_{ij} 
- {\ell \over 2 \, u(\alpha)} \, \CG_{ij}
\right]
\, .
\label{by}
\end{eqnarray}
Here $\theta_{ij}$ is the extrinsic curvature of the fixed-$r$
surface, $\theta$ is the trace of $\theta_{ij}$, and  $\CG_{ij}$ is
the Einstein tensor constructed from the boundary metric.   From the
Hamilton-Jacobi perspective, $\tau_{ij}$ is simply the variable
conjugate to the boundary metric $\gamma_{ij}$, given the action
(\ref{bulkact}) as a functional of boundary data~\cite{brownyork}.

For solutions of the form (\ref{warpmet}), the  trace of $\tau_{ij}$
is
\begin{equation}
\tau \equiv \gamma^{ij} \, \tau_{ij}  = 
{3 \over 2 \pi G_N \ell} \, \left[
\ell \, {d A \over d r} 
- u(\alpha)
\right]
\, .
\label{tau}
\end{equation}
We found solutions to the equations of motion by assuming $d \alpha^I
/ dr \neq 0$, so that $A'$ and $\alpha^{I'}$ could be expressed as
functions of $\alpha^I$.  Under these conditions, the trace of the
Hamilton-Jacobi expression for the Brown-York stress tensor is
proportional  to
\begin{equation}
{\delta S_h \over \delta A} = {\delta S_h \over \delta \alpha^I} \,
{\delta \alpha^I \over \delta A} 
\, .
\label{tracerg}
\end{equation}
(Recall that $\tau = (D/\sqrt{-\gamma}) \, (\delta S/\delta
A)$).

In the connection between CFTs and gravity on AdS spaces, position
in the radial direction is known to play the role of the scale in 
dual field theory computations.   It is difficult to map the
radial coordinate directly into field theory scales because of the
possibility of radial reparametrization.  However, there is an
invariant way to study radial positions in classical solutions of the
form (\ref{warpmet}) -- we will associate the warp factor in the
metric ($A$), with the $\log$ of $\lambda$, the  scale in the
definition of the beta functions in (\ref{trace}).   The map between
the field theory RG equations and the gravitational equations of
motion is then:
\begin{center}
\begin{tabular}{|c|c|c|c|} \hline
 $G^{IJ} \equiv 6 \, g^{IJ}$    &  $\phi^I \equiv \alpha^I$         &  
          $a \equiv {2\,\pi\,G_N\, \ell \over 3}  $     
&
$\log\lambda = A$
\\
\hline 
$\langle T \rangle \equiv \tau$ & $V(\phi) \equiv -v(\alpha)$ &
          $U(\phi) \equiv u(\alpha)$
&
 
\\ 
\hline
\end{tabular}
\end{center}
Using these substitutions and the fact that $A' = (a\, \tau +
u(\alpha))/\ell$ (\ref{tau}), it easy to show that the equations of motion
(\ref{alphasoln}) and (\ref{secondeom}) are exactly the gradient beta
function equation (\ref{betaT}) and the potential equation
(\ref{potential}) respectively.  As discussed, simultaneous solution of
(\ref{alphasoln}) and (\ref{secondeom}) is sufficient to solve the complete
coupled second-order equations for the scalars and the spacetime metric
subject to the 4d Poincar\'e invariant ansatz (\ref{warpmet}).  Identifying
variations of $\Gamma$, the QFT effective action as a function of
couplings, with variations of $S$, the spacetime action as a function of
boundary data, the expression for the trace of Brown-York stress tensor
(\ref{tracerg}) becomes equivalent to the field theory RG equation
(\ref{trace}).

We have mapped the boundary values of the spacetime scalar fields and
the metric to field theory couplings.  That accounts for half the
integration constants of the equations of motion (\ref{firsteom}) --
(\ref{scaleom}).  The remaining constants are the radial derivatives
of bulk fields which, in the AdS/CFT context, were related to
expectation values of operators in the CFT~\cite{bklt}.  A similar
relationship holds here; for example, $\langle T \rangle = \tau \sim
dA/dr$ according to (\ref{tau}).  Our analysis treats RG flow in the
vacuum state where operator expectation values have been set to zero.
From the gravitational perspective, this should correspond to fixing
the radial derivatives of bulk fields to yield the lowest energy given
the boundary values.

In perturbative string theory, a central consistency condition is the
requirement of world-sheet conformal invariance, which implies that strings
can consistently propagate only in backgrounds which satisfy the string
equation of motion. In other words, the condition that the beta function is
zero is equivalent to the target space equation of
motion~\cite{stringbooks}.  Equation (\ref{betaT}) is an analogue of this
statement in the context of holographic duality.  Supplemented by the
requirement (\ref{potential}) on the norm of the beta functions,
(\ref{betaT}) yields the full equations of motion of the spacetime fields.
As we shall see, the potential in the gradient beta function equation is
intimately related to a c-function which decreases monotonically along our
RG flows.  After the identifications above, (\ref{potential}) became the
condition of $r$-reparametrization invariance of the 5-dimensional
spacetime. This suggests that it should be directly related to invariance
of the field theory under redefinitions of the floating scale.

\vskip0.1in
{\leftline {\bf {\large C-functions and an analogue Zamolodchikov metric}}}

Any field theory satisfying the constraint (\ref{betaT}) has a c-function
that decreases monotonically during RG-flow.  The basic reason for this is
that $\langle T \rangle$ is identified with the trace of $\tau_{ij}$, the
quasilocal stress tensor of spacetime given in (\ref{by}).  Since the
intrinsic curvatures of the boundaries at fixed $r$ vanish for metrics of
the form (\ref{warpmet}), the trace of the stress tensor depends only on
the extrinsic curvature:
\begin{equation}
\langle T \rangle \equiv \tau = -{3 \over 8\pi G} \, \left( \theta +
{4 \, u(\alpha) \over \ell} \right) \, .
\end{equation}
The Raychaudhuri equation  implies a monotonic radial flow of $\theta$,
\begin{equation}
{d \theta \over d r} 
\leq 0 \, , 
\label{monoton}
\end{equation}
so long as a form of the weak positive energy condition is satisfied
by the scalar fields~\cite{wald}.\footnote{Monotonicity follows from
the Raychaudhuri equation describing evolution of the expansion
$\theta$ along null curves, applied to static metrics of the form
(\ref{warpmet}).  If ${\CT}_{ab}$ is the matter stress tensor and
$\CT$ is its trace, the required energy condition is that ${\CT}_{ab}
\, k^a \, k^b \geq 0$ for all null $k^a$.  In~\cite{wald}, the weak
energy condition is defined with respect to timelike $\xi^a$:
${\CT}_{ab} \, \xi^a \, \xi^b \geq 0$.  Both this, and the strong
positive energy condition in~\cite{wald}, ${\CT}_{ab} \, \xi^a \,
\xi^b \geq -(1/2)\, \CT$, imply the positivity of ${\CT}_{ab} \, k^a
\, k^b$.}

This fact, coupled with dimensional analysis and Bousso's covariant entropy
formula~\cite{bousso}, has led Sahakian~\cite{vatche} to propose a
candidate gravitational analogue of a c-function:\footnote{We are
specializing the proposal to metrics of the form (\ref{warpmet}).}
\begin{equation}
c \propto {1 \over G_N \, \theta^3} 
\label{cprop}
\end{equation}
(See~\cite{alvgom,adsrgsolns,italians,firstorder} for interesting examples and related
proposals.)  Translating into field theory variables, the candidate
c-function is
\begin{equation}
c =
c_0 \, \left( a \, \langle T \rangle + U(\phi) \right)^{-3}  \, ~~~~~;~~~~~
c_0 = {b \, \ell^3 \over G_N}
\label{ccand}
\end{equation}
with $b$ a dimensionless numerical constant.  The Raychaudhuri
equation automatically implies monotonicity of this c-function.
Accepting this proposal, the gradient beta function equation
(\ref{betaT}) may be rewritten as
\begin{equation}
\beta^I = 
{1\over 3} \, G^{IJ} \, {\partial \log c \over \partial \phi^J} 
\equiv 
\tilde{G}^{IJ} \, {\partial c \over \partial \phi^J} 
\,  .
\label{betac1}
\end{equation}
This formula gives a useful consistency check on the identifications
we have been performing between field theory and gravitational
quantities.  We will show that 
$\tilde{G}^{IJ} = G^{IJ}/3c$ 
determines the normalization of the field theory 2-point function in the
vicinity of  a conformal point.

To see this, recall that perturbing around a conformal point by some
marginal operators $O_I$, an analogue of the c-function in two dimensions
can be defined as
\begin{equation}
T(x) \propto \beta^{I} \, O_I ~~~~~;~~~~~
\langle T(k) \, T(-k) \rangle = \tilde{c} \, k^{4} 
\label{analog}
\end{equation}
in terms of the trace of the stress tensor $T = T^i_i$.  At a conformal
point, the 2-point function of $O_I$ is given by $\langle O_I(x) \, O_J (0)
\rangle = {G^Z_{IJ} \over x^8 }$, so that in the vicinity of this point
\begin{equation}
\langle O_I(k) O_J (-k) \rangle = k^4 \, \left( {\rm const.} + {G^Z_{IJ}} \,  \log{{k \over
\Lambda_0}} + \cdots \right)  \equiv
G^Z_{IJ} \, f(k) + \cdots
\label{k4}
\end{equation}
Here $G^Z_{IJ}$ is an analogue of the Zamolodchikov metric defined in two
dimensions.  The last two equations together imply that 
\begin{equation}
\tilde{c} = {\rm const.} + \beta^I \beta^J
{G^Z_{IJ}} \log{{k \over \Lambda_0}} + ... 
\end{equation}
An alternative expression for $c$  can be obtained by expanding the
couplings around the conformal point as  $\phi^I= \phi^I_{0} + \beta^I
\log{{k \over \Lambda_0}}+ \cdots$ and expanding $c(\phi)$ around $\phi_0$:
\begin{equation}
\tilde{c} = {\rm const.} + { \partial c \over \partial \phi^I} \, \beta^I \,
\log{{k \over \Lambda_0}} + \cdots
\end{equation}
The two expressions for $\tilde{c}$ are equal close to a conformal point if
\begin{equation}
\beta^I = G_Z^{IJ} {\partial \tilde{c} \over \partial \phi^J} \, ,
\label{betac2}
\end{equation}
$G^{IJ}_Z$ being the inverse of $G^Z_{IJ}$.  In two dimensions, this
formula is exactly true by Zamolodchikov's theorem~\cite{zamibaby}, while
the above is an approximate derivation in four dimensions, in the vicinity
of a conformal point.
In fact, various candidates have been proposed for a c-function in
4-dimensional field theory~\cite{cardy}.  The coefficient $\tilde{c}$ in
(\ref{analog}) is one of these, and is related to the coefficient of the
Weyl tensor squared term in the conformal anomaly~\cite{crefs0,crefs1}.  Another
candidate is the coefficient of the Euler invariant in the anomaly.  The
material point for us is that equation (\ref{betac2}) holds true near a
conformal point for either candidate c-function
(see~\cite{crefs0,crefs1,crefs2} and  references therein).

Comparing (\ref{betac1}) and (\ref{betac2}) shows that consistency of the
identification of the c-function in the former requires that
$\tilde{G}_{IJ}$ determine the 2-point function of $O_I$ at the conformal
point. At the UV conformal point, the beta functions and $\langle T
\rangle$ are zero, while $u(\phi) = 1$; we then expect that 
$\langle O_I(k) O_J(-k) \rangle = f(k) \, \tilde{G}_{IJ} = 3 c\, f(k) \,
G_{IJ} $.  This serves as a consistency check on the dictionary between
field theory and gravity quantities that we are developing.  We equated
the spacetime action $S$, as a functional of boundary data, to the field
theory quantum effective action $\Gamma$, as a functional of couplings.  So
the field theory 2-point function relates directly to the spacetime
propagator between boundary points for the scalars $\alpha_I$\footnote{Note
the lower index.}.  We would like to see that this is proportional to $f(k)
\, G_{IJ} \, (G_N / \ell^3) $.

At the ultraviolet conformal point, the potential $v(\alpha)$ in
(\ref{bulkact}) was normalized to $-1$, and the equations of motion are
solved to give pure anti-de Sitter space.  So the scalar propagator in
spacetime is given by the standard computation in the AdS/CFT
correspondence~\cite{juanads,gkpw,freedman} and is proportional in momentum
space to $f(k) = k^4 \log k$.  The normalization of the scalar fields in
(\ref{bulkact}) implies that the scalar propagator for $\alpha_I$ is
proportional to $G_N \, g_{IJ}$.  Since $G_N$ has length dimension three
and $\ell$ is only remaning length scale, the scalar propagator must yield,
via the QFT-gravity dictionary,
\begin{equation}
\langle O_I\, O_J \rangle \propto f(k) \, G_{IJ} \, {G_N \over \ell^3}
+ \cdots  
\end{equation}
in the vicinity of the UV conformal point.  In the identification of
the RG equations and the spacetime equations of motion, only the
combination of parameters $a = 2\pi G_N \ell/3$ appeared.  Now we see
that this data can combine with the normalization of the 2-point
function of $O_I$ to separately yield the Newton constant ($G_N$) and
the spacetime curvature scale ($\ell$).  Analysis of higher point
functions would give further consistency conditions.

The results we have accumulated suffice to show that the RG flow of any
theory with a monotonic c-function can be be rewritten in terms of a 5d
gravity background under some conditions.  Suppose we are given as
data the beta functions, stress tensor and monotonic c-function of a
4-dimensional QFT.  The relation (\ref{betac1}) serves to define a metric
$G^{IJ}$, which we require to be positive and symmetric.  Equation
(\ref{ccand}) defines the potential $U(\phi)$, and thereby the gradient
beta function equation (\ref{betaT}).  Finally, consider the scale
variation of the c-function:
\begin{equation}
{d\log c \over  d \log \lambda} =
\beta^I \, {\partial \log c \over \partial \phi^I}
\label{cev1}
\end{equation}
Inverting
(\ref{betac1}) to get an equation for $\partial\log  c/\partial\phi^J$, we
find:
\begin{equation}
{d\log c \over  d \log \lambda} \propto
- G_{IJ} \, \beta^I \, \beta^J
\label{cev2}
\end{equation}
The right hand side of this equation defines the potential $V(\phi)$ in
(\ref{potential}).  We now see that $V(\phi)$ is related to the deviation
of the c-function away from its value at the conformal point.  In summary,
given the beta functions, stress tensor and monotonic c-function of a 4d
QFT, we have defined a metric and two potentials.  As we have shown, these
are the elements of a five-dimensional gravitational Lagrangian whose
equations of motion reproduce the field theory RG equations.

\vskip0.1in
{\leftline {\bf {\large Discussion}}}

The Hamiltonian constraint (\ref{secondeom}) expresses radial
reparametrization invariance of solutions to the action (\ref{bulkact}).
If radial positions are mapped to field theory scales, we would expect to
extract a ``holographic'' RG equation from this constraint.  In \cite{dbvv}
de Boer, Verlinde and Verlinde achieve this goal by separating the
five-dimensional bulk spacetime action $S_h$ bounded at a given radial
position into one piece which is local in the boundary data ($S_{{\rm
l}}$), and another which is non-local ($S_{{\rm nl}}$).  The Hamiltonian
constraint, i.e. the Hamilton-Jacobi equation for the total spacetime
action $S_h$, is rewritten as a first order RG equation of the
Callan-Symanzik type for the non-local action $S_{{\rm nl}}$.  The
beta-functions are implicitly determined by the form of the local action
$S_{{\rm l}}$ since they are defined as ratios of ${\delta S_{{\rm l}}
\over \delta{A}}$ and ${\delta S_{{\rm l}} \over \delta \phi^I}$.  This
definition agrees with the beta function derived from the conformal Ward
identity, i.e., the trace of the Brown-York stress tensor. Therefore,
knowledge of $S_{{\rm l}}$ completely specifies the form of the beta
function.  Combined with the original Hamiltonian constraint, this
determines all the equations of motion, as we have shown above.
Furthermore, when the space is $3+1$ Poincar\'e invariant, the local action
$S_{{\rm l}}$ is determined by the vacuum energy density of the boundary
field theory, which is proportional to the trace of the stress-energy
tensor. We have derived an RG equation from gravity by directly computing
the trace of the quasilocal stress tensor. The resulting beta function
agrees exactly with the one discussed in~\cite{dbvv}.  The Hamiltonian
constraint (\ref{secondeom}) can be rewritten as (\ref{potential}) which,
as we have seen (\ref{cev2}), simply defines the flow of the $c$ function
in our formalism.

We have shown that if the beta functions of a 4d field theory are given by
gradients of a certain potential, then the RG flow can be expressed as a
classical solution of 5 dimensional gravity coupled to scalar
fields.\footnote{By construction, the examples of holographic RG flows
developed in~\cite{MIT,adsrgsolns,italians,firstorder} all fit our
formalism.}  The Raychaudhuri equation in 5 dimensions automatically
guaranteed a c-theorem for such flows.  Equivalently, a 4d RG flow relates
to 5d gravity when the beta functions are given by the gradient of a
monotonic c-function (\ref{betac1}), while the c-function is related to the
trace of the stress tensor as in (\ref{ccand}).  Given the beta functions,
stress tensor and c-function of a field theory we can always rewrite RG flow
in terms of 5d gravity so long as the symmetric, positive metric $G^{IJ}$
in (\ref{betac1}) can be defined.  Similar results may be derived for RG
flows of field theories in other dimensions.  Famously, beta functions of
{\it any} renormalizable theory in two dimensions can be expressed as
gradients of a c-function~\cite{zamibaby} with a positive symmetric metric
appearing in (\ref{betac1}).
It has been suggested that a holographic c-function appropriate to
2-dimensional theories will satisfy $c \sim 1/(a\langle T \rangle +
U(\phi))$~\cite{vatche}.  Using this to define $U(\phi)$, and
(\ref{potential}) to define $V(\phi)$, we expect that any such RG flow is
expressible in terms of a gravity background.

What characterizes a theory which realizes gradient beta functions of the
form discussed in this paper?  By construction, the large $N$, conformal
theories appearing in the AdS/CFT correspondence have the requisite traits.
However, the symmetries and properties guaranteeing (\ref{betaT}) (or a
monotonic c-function) have not been generally understood.  Based on the
AdS/CFT experience, we expect that a theory might have an RG flow
satisfying (\ref{betaT}) in some limit of parameters, but that deviations
from the limit produce systematic corrections.  These should be compared to
modified spacetime equations of motion arising from higher derivative terms
added to the action (\ref{bulkact}).  In string theory, such terms
originate in propagation of the excited states of string and in loop
corrections.

Only some field theory RG schemes can be expected to have holographic
descriptions in gravity.  Within our analysis, there is a scheme dependence
in the definition of the stress tensor $T$ and the potentials $U$ and $V$
in field theory.  However, these stress tensor ambiguities can be precisely
matched by modifications in the counterterm scheme (\ref{counter}) for the
gravitational stress tensor.\footnote{One such ambiguity, affecting the
definition of the trace anomaly on a curved manifold is discussed
in~\cite{adsstress}.}  More generally, different RG schemes involve
different ways of imposing a cutoff and integrating out modes or,
alternatively, different choices of counterterms.  The AdS/CFT
correspondence suggests that an appropriate class of RG schemes coarsens
field variables by convolving them against a smearing kernel~\cite{adsrg}.
Even this cannot be entirely sufficient as the field theory approaches the
deep infrared.  In this limit, the gravity description involves a large
bubble of essentially flat space at the center of an AdS spacetime.  Since
the field theory is in the deep infrared, it appears that the physics of
homogeneous modes describes the flat space region.  In the AdS/CFT case,
these modes constitute the quantum mechanics of a large matrix, in an echo
of the M(atrix) model of M-theory~\cite{bfss}.  This suggests that recovery
of the equations of motion of a flat space region in AdS requires
implementation of a ``matrix renormalization group'' relating the physics
of $SU(N)$ to $SU(N-k)$.

Holographic realizations of the renormalization group associate field
theory scales with radial positions in a higher dimensional spacetime.  The
full set of 5-dimensional diffeomorphisms can put ``bumps'' in surfaces at
fixed radial positions. Recovering such transformations from 4-dimensional
field theory will certainly involve a {\it local} notion of the
renormalization group where the coarsening scale varies from point to
point.  This automatically requires consideration of theories with
spatially varying couplings.  It will be illuminating to elucidate the
relation between local RG invariance in field theory and higher dimensional
diffeomorphism invariance.

\vskip0.1in {\leftline {\bf Acknowledgements}}

We have benefitted from discussions with M.~Douglas, P.~Ho\v{r}ava,
P.~Kraus, S.~Shenker, E.~Verlinde, H.~Verlinde, N.~Warner, E.~Witten, and,
particularly, J.~de Boer and R.~Gopakumar.  {\small V.B.}  was supported by
the Harvard Society of Fellows, the Milton Fund of Harvard University, and
NSF grant NSF-PHY-9802709. {\small E.G.} was supported by DOE grant
DE-FG03-92ER40701.  {\small D.M.} was supported by DOE grant
DE-FG03-84ER40168.  {\small V.B.} is grateful to the University of Chicago
and UCLA where part of this work was completed. {\small D.M.} enjoyed the
hospitality of the University of Illinois at Chicago during this project.

Recent work which overlaps with the content of this paper appears in
\cite{adsrgothers}.


\end{document}